\documentclass[aps,pra,twocolumn,showpacs,amsmath,amssymb,longbibliography]{revtex4-1}

\usepackage{graphicx}
\usepackage{dcolumn}
\usepackage{bm}

\begin{document}

\title{Heralded magnetism in non-Hermitian atomic systems}

\author{Tony E. Lee}
\affiliation{ITAMP, Harvard-Smithsonian Center for Astrophysics, Cambridge, MA 02138, USA}
\affiliation{Department of Physics, Harvard University, Cambridge, MA 02138, USA}
\author{Ching-Kit Chan}
\affiliation{ITAMP, Harvard-Smithsonian Center for Astrophysics, Cambridge, MA 02138, USA}
\affiliation{Department of Physics, Harvard University, Cambridge, MA 02138, USA}

\date{\today}

\begin{abstract} 
Quantum phase transitions are usually studied in terms of Hermitian Hamiltonians. However, cold-atom experiments are intrinsically non-Hermitian due to spontaneous decay. Here, we show that non-Hermitian systems exhibit quantum phase transitions that are beyond the paradigm of Hermitian physics. We consider the non-Hermitian XY model, which can be implemented using three-level atoms with spontaneous decay. We exactly solve the model in one dimension and show that there is a quantum phase transition from short-range order to quasi-long-range order despite the absence of a continuous symmetry in the Hamiltonian. The ordered phase has a frustrated spin pattern. The critical exponent $\nu$ can be 1 or 1/2. Our results can be seen experimentally with trapped ions, cavity QED, and atoms in optical lattices.
\end{abstract}

\maketitle

\section{Introduction}

A quantum phase transition occurs when the ground state of a many-body system experiences a sudden change as a parameter is tuned through a critical point \cite{sachdev11}. In the last few decades, a great deal of work has led to a broad understanding of quantum criticality. It is now known how the ordering and critical exponents depend on the symmetries of the Hamiltonian. In the study of quantum phase transitions, it is generally assumed that the Hamiltonian is Hermitian, since solid-state systems are governed by Hermitian Hamiltonians. However, it turns out that non-Hermitian Hamiltonians arise in cold-atom experiments due to spontaneous decay \cite{dalibard92,molmer93,dum92,wiseman96,plenio98}. Thus, cold-atom experiments provide the opportunity to discover new classes of phase transitions beyond the framework of Hermitian critical phenomena.

In this paper, we exactly solve the non-Hermitian $XY$ model in one dimension and show that it violates several tenets of Hermitian systems. There is a sharp transition already for two atoms. In a long chain, there is a quantum phase transition from short-range order to quasi-long-range order despite the absence of a continuous symmetry in the Hamiltonian. The ordered phase has a frustrated spin pattern despite the short-range interaction. The critical exponent $\nu$ can be either 1 or 1/2, the latter value being unusual for a spin chain. The phase boundaries are also completely modified. 

\begin{figure}[t]
\centering
\includegraphics[width=3.2 in,trim=0in 0in 0in 0in,clip]{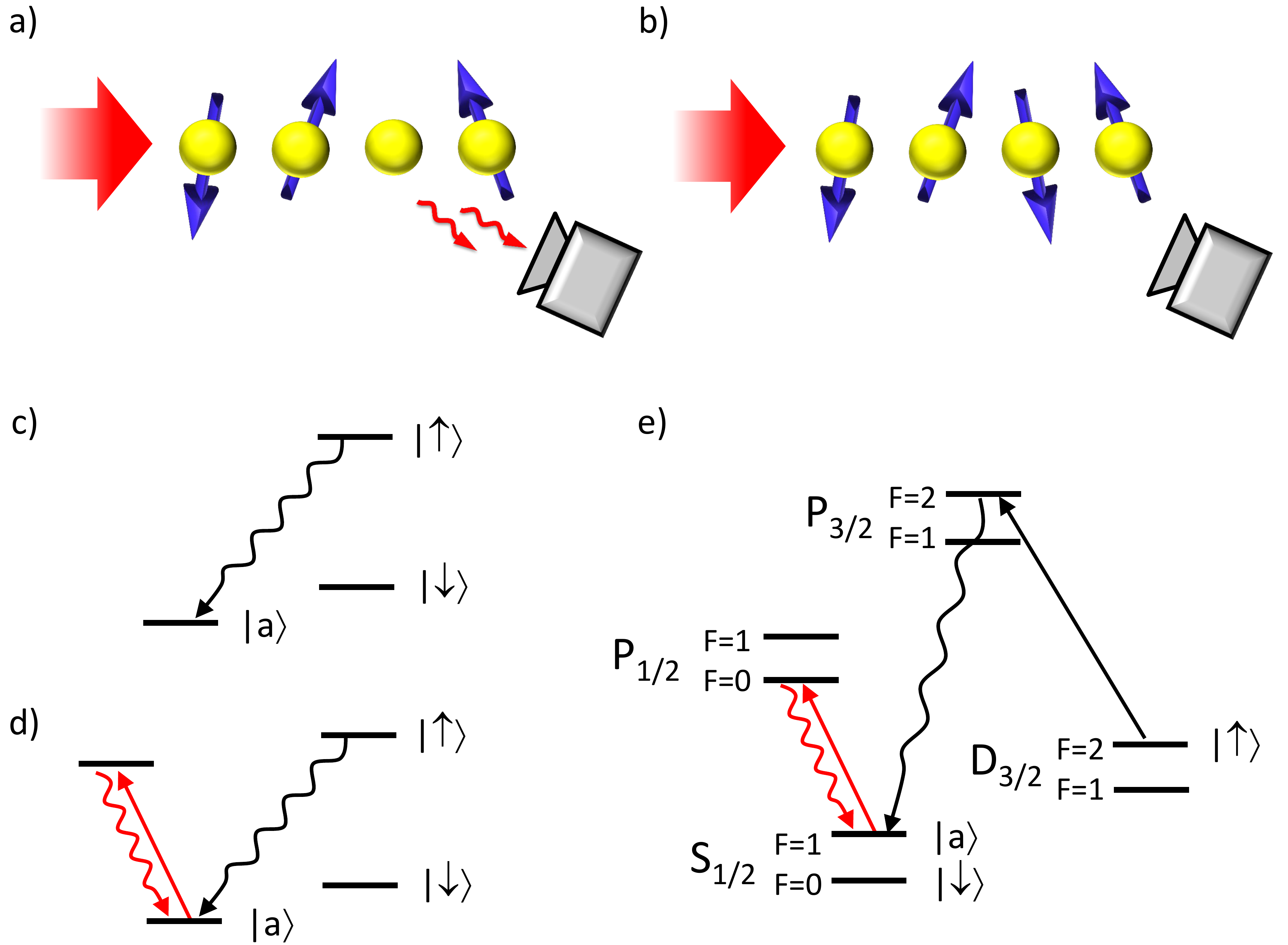}
\caption{\label{fig:level_scheme}(a) Experimental setup with a chain of atoms. The population in the auxilary state $|a\rangle$ is measured by scattering photons off of it. (b) The non-Hermitian model is heralded by the absence of population in $|a\rangle$, i.e., the absence of fluorescence. (c) $\left|\uparrow\right\rangle$ decays into the auxiliary state $|a\rangle$. (d) The population of $|a\rangle$ is measured by exciting it with a laser and detecting the fluorescence (red arrows). (e) Level scheme for an ${}^{171}\text{Yb}^+$ ion, showing optical pumping (black arrows) and detection (red arrows).}
\end{figure}

As we discuss in detail below, the non-Hermitian $XY$ model can be experimentally simulated using three-level atoms in a variety of setups, including trapped ions, cavity QED, and atoms in optical lattices. The non-Hermiticity is due to measuring whether a spontaneous decay has occured [Fig.~\ref{fig:level_scheme}(a--b)]. The non-Hermitian model is heralded by the absence of a spontaneous decay event, which can be measured with a high degree of accuracy \cite{katz06,myerson08,sherman13}. This is similar to heralded entanglement protocols in which a measurement signals the preparation of the desired state without destroying it \cite{duan01,chou05,matsukevich06,moehring07,casabone13}. In fact, for realistic experimental parameters, one can implement the non-Hermitian $XY$ model with at least $20$ atoms with a higher success probability than heralded entanglement protocols. We also show that 20 atoms are enough to experimentally observe our results.

In recent years, non-Hermitian models have drawn interest because they exhibit a variety of rich behavior \cite{moiseyev11,berry98,heiss12,liertzer12}, such as localization \cite{hatano96,refael06}, $\mathcal{PT}$ symmetry \cite{bender98,giorgi10,zhang13}, and spatial condensate order \cite{otterbach14}. Recent works have discovered dynamical phase transitions that occur when physical parameters are extended to the complex plane \cite{wei12,heyl13,wei14,peng14,hickey13}; these works motivate the study of correlation functions as well as the experimental implementation of non-Hermitian Hamiltonians. In this paper, we show how non-Hermitian quantum mechanics leads to new magnetic behavior that can be observed in current cold-atom setups.


\section{Model}

Consider a one-dimensional chain of atoms that interact via the anisotropic $XY$ model with nearest neighbor interactions \cite{lieb61,barouch71,sachdev11}:
\begin{eqnarray}
H&=&\sum_n (J_x\sigma^x_n\sigma^x_{n+1} + J_y\sigma^y_n\sigma^y_{n+1}). \label{eq:H}
\end{eqnarray}
Each atom has states $\left|\uparrow\right\rangle$ and $\left|\downarrow\right\rangle$, which constitute the relevant Hilbert space, and $\sigma^x_n,\sigma^y_n,\sigma^z_n$ are the Pauli matrices for atom $n$. We assume that $\left|\uparrow\right\rangle$ decays into an auxiliary state $|a\rangle$ with rate $\gamma$ [Fig.~\ref{fig:level_scheme}(c)]. Then in the absence of a decay event, the system evolves with the effective non-Hermitian Hamiltonian \cite{dalibard92,molmer93,dum92,wiseman96,plenio98},
\begin{eqnarray}
H_{\text{eff}}&=&H-\frac{i\gamma}{4}\sum_n (\sigma_n^z + 1), \label{eq:Heff}
\end{eqnarray}
as explained in Appendix ~\ref{sec:app_nonh}. This counter-intuitive effect is due to the fact that the atoms are coupled to the environment, and the environment continuously measures whether a photon has been emitted. Even when no atom has decayed to $|a\rangle$, the null measurement of photons still affects the wavefunction of the atoms. The measurement back-action is accounted for by the non-Hermitian term in Eq.~\eqref{eq:Heff}, which transfers population from $\left|\uparrow\right\rangle$  to $\left|\downarrow\right\rangle$ in a non-unitary way. (When calculating observables, the wavefunction should be normalized to 1.) The non-Hermitian evolution has been experimentally observed for a single atom \cite{katz06,sherman13}.

\begin{figure}[t]
\centering
\includegraphics[width=3.5 in,trim=1.in 4.1in 1in 4.1in,clip]{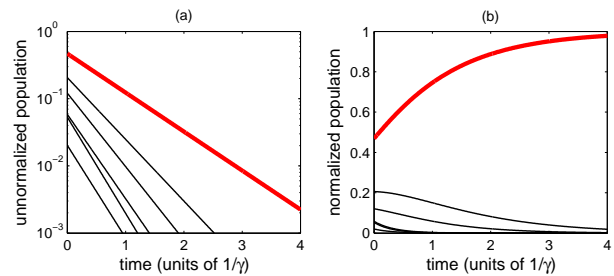}
\caption{\label{fig:population}Population in the six slowest-decaying eigenstates of $H_\text{eff}$ during the non-Hermitian evolution (a) without normalization and (b) with normalization. This is from exact diagonalization of $N=10$ atoms with open boundary conditions, initial condition $\left|\downarrow\downarrow\cdots\downarrow\right\rangle$, $J'=0.12\gamma$, and $J=0$. The thick red line denotes the steady state. }
\end{figure}

A many-body wavefunction can be written as a superposition of the eigenstates of $H_{\text{eff}}$. The eigenvalues of $H_{\text{eff}}$ are complex and have negative imaginary parts \cite{amir08}.  When the wavefunction is evolved using $\exp(-iH_{\text{eff}}\,t)$, the imaginary parts of the eigenvalues cause the weight in each eigenstate to decrease over time [Fig.~\ref{fig:population}(a)]. However, the eigenvalue with the largest (least negative) imaginary part decreases the slowest, so after a sufficient amount of time, the wavefunction consists mostly of the corresponding eigenstate [Fig.~\ref{fig:population}(b)]. We call this surviving eigenstate the steady-state wavefunction, and we study its phase diagram.


The experimental protocol is as follows. In each experimental run, the atoms interact via Eq.~\eqref{eq:H} while possibly decaying into $|a\rangle$. After a sufficient amount of time, one measures the population in $|a\rangle$ to check whether any atom has decayed. The experimental runs without decay events evolve solely with $H_{\text{eff}}$ and thus simulate the model of interest. To measure the population in $|a\rangle$, one laser-excites $|a\rangle$ and looks for fluorescence [Fig.~\ref{fig:level_scheme}(d)]. If an atom is in $|a\rangle$, it will scatter many photons during this measurement; the absence of fluorescence signals the absence of population in $|a\rangle$. This way, one can measure the population with almost 100\% efficiency \cite{myerson08}. If one finds no population in $|a\rangle$, then the atoms are in the steady state wavefunction, and one proceeds to measure the correlation functions discussed in Sec.~\ref{sec:long}. (If the steady state is difficult to reach, one can use Fourier analysis to distinguish between eigenstates, as explained in Sec.~\ref{sec:experiment}).

This scheme is similar to heralded entanglement protocols: when a trial succeeds, it is heralded by a measurement outcome \cite{duan01,chou05,matsukevich06,moehring07,casabone13}. Importantly, the heralding does not itself destroy the state. The magnetic model we study is heralded by the absence of population in $|a\rangle$. In Sec.~\ref{sec:experiment}, we show that one can implement this scheme with at least $N=20$ atoms with a relatively high success probability.


We emphasize that our model is different from the typical dissipative model described by a master equation for the density matrix $\rho$,
\begin{eqnarray}
\dot{\rho}&=&-i[H,\rho] + \gamma\sum_n\left[\sigma^-_n\rho\sigma^+_n -\frac{1}{2}(\sigma^+_n\sigma^-_n\rho + \rho\sigma^+_n\sigma^-_n)\right].\nonumber\\ \label{eq:master}
\end{eqnarray}
This master equation applies when $\left|\uparrow\right\rangle$ decays into $\left|\downarrow\right\rangle$ and was discussed previously in Refs.~\cite{lee13c,joshi13}. Below, we show that the phase diagram of $H_{\text{eff}}$ is quite different from that of the master equation as well as $H$. The master equation with dissipation on the boundaries was studied in Refs.~\cite{prosen08,banchi14}. Master equations of other spin models were considered in Refs.~\cite{sieberer13,nissen12,carr13a,malossi14,lee11,lee12,hu13,qian13,fossfeig13,morrison08,bardyn12,lemeshko13,honing12,horstmann13}. We also emphasize that our model is different from the spin-boson model \cite{leggett87}: we study the nonequilibrium steady state of the system, instead of the equilibrium ground state of the system and environment.


For convenience, we rewrite $H_{\text{eff}}$ as
\begin{eqnarray}
H_{\text{eff}}&=&\sum_n \Big[2J(\sigma^+_n\sigma^-_{n+1} + \sigma^-_n\sigma^+_{n+1}) \nonumber\\
&&\quad\quad + 2J'(\sigma^+_n\sigma^+_{n+1} + \sigma^-_n\sigma^-_{n+1}) - \frac{i\gamma}{4}(\sigma_n^z + 1)\Big], \quad \label{eq:Heff2}
\end{eqnarray}
where $J=(J_x+J_y)/2$ and $J'=(J_x-J_y)/2$. There is competition between the non-Hermitian term (measured by $\gamma$) and the anisotropic interaction (measured by $J'$) that coherently excites pairs of atoms. This competition leads to the critical behavior discussed below.

\section{Two atoms}\label{sec:N2}

\subsection{Exceptional point}

We first consider the case of two atoms, since it is the easiest to realize experimentally. (We assume periodic boundary conditions to match up with the results for larger chains.) $H_{\text{eff}}$ has four eigenvalues: $\lambda_1^{\pm}=-i\gamma/2 \pm (1/2)\sqrt{64J'^2-\gamma^2}$ and $\lambda_2^{\pm}=-i\gamma/2 \pm 4J$, corresponding to eigenstates,
\begin{eqnarray}
\left|u_1^{\pm}\right\rangle=\frac{1}{\mathcal{N}}\left(
\begin{array}{c}
\frac{-i\gamma\pm\sqrt{64J'^2 - \gamma^2}}{8J'} \\ 0 \\ 0 \\ 1
\end{array}
\right), \quad
\left|u_2^{\pm}\right\rangle=\left(
\begin{array}{c}
0 \\  \frac{\pm1}{\sqrt{2}} \\ \frac{1}{\sqrt{2}} \\ 0
\end{array}
\right),\quad \label{eq:eigenstates}
\end{eqnarray}
in the basis $\{\left|\uparrow\uparrow\right\rangle,\left|\uparrow\downarrow\right\rangle,\left|\downarrow\uparrow\right\rangle,\left|\downarrow\downarrow\right\rangle\}$. $\mathcal{N}$ is a normalization constant.

When $|J'|<\gamma/8$, the steady state is $\left|u_1^+\right\rangle$ since its eigenvalue $\lambda_1^+$ has the largest imaginary part. A degeneracy occurs when $|J'|= \gamma/8$ since $\lambda_1^+=\lambda_1^-$ there. This degeneracy leads to nonanalytic behavior of the steady state as $|J'|$ passes through $\gamma/8$. The nonanalyticity is most easily seen in $\langle \sigma^z_n\rangle$: when $|J'|<\gamma/8$, $\langle \sigma^z_n\rangle=-\sqrt{1-64(J'/\gamma)^2}$, while when $|J'|\geq\gamma/8$, $\langle \sigma^z_n\rangle=0$ as plotted in Fig.~\ref{fig:sz}(a).  


Thus, the non-Hermitian model has a sharp transition already for two atoms. The ability to have sharp transitions in \emph{finite} systems is a unique feature of non-Hermitian Hamiltonians, and non-Hermitian degeneracies are known as ``exceptional points'' \cite{moiseyev11,berry98,heiss12}. In contrast, Hermitian Hamiltonians exhibit avoided crossings in finite systems.




\subsection{Physical interpretation}

Suppose the wavefunction starts out as a superposition of eigenstates of $H_{\text{eff}}$. Under the non-Hermitian evolution (and before normalizing the wavefunction), the population in each eigenstate decreases over time due to the chance of a decay event to $|a\rangle$. The rate of this decrease for each eigenstate is
\begin{eqnarray}
-2\,\text{Im }\lambda&=&N\gamma\left(\frac{\langle \sigma^z_n\rangle+1}{2}\right), \label{eq:decay_rate}
\end{eqnarray}
as can be checked using Eq.~\eqref{eq:eigenstates}. There is thus a close connection between an eigenvalue's imaginary part and $\langle\sigma^z_n\rangle$.

Since the steady state is the eigenstate whose eigenvalue has the largest (least negative) imaginary part, it is also the eigenstate with the smallest $\langle \sigma^z_n\rangle$. This is because the smaller $\langle \sigma^z_n\rangle$ is, the smaller the probability of a decay event to $|a\rangle$. If there is no decay event for a sufficient amount of time, the wavefunction consists only of the steady state, because the null measurement of a decay event projects the system into the eigenstate least likely to have a decay event.

\begin{figure}[t]
\centering
\includegraphics[width=3.5 in,trim=1.2in 4in 0.9in 4.2in,clip]{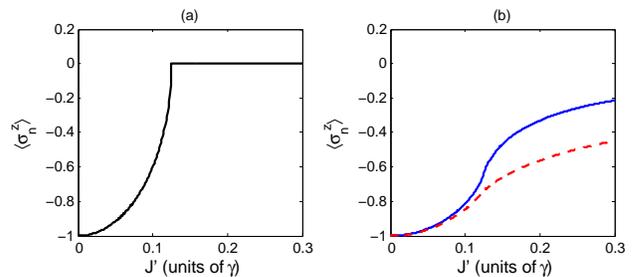}
\caption{\label{fig:sz}$\langle\sigma^z_n\rangle$ for (a) two atoms and (b) infinite chain with $J=0$ (blue solid line) and $J=0.1\gamma$ (red dashed line). The critical point is $J'=\gamma/8$ for all cases. Panel (a) is independent of $J$.}
\end{figure}

$J'$ is a coherent process that excites pairs of atoms and thus acts as a drive [Eq.~\eqref{eq:Heff2}]. As $J'$ increases, $\langle \sigma^z_n\rangle$ of the steady state increases [Fig.~\ref{fig:sz}(a)], and so its eigenvalue's imaginary part decreases (becomes more negative) [Eq.~\eqref{eq:decay_rate}]. In other words, as $J'$ increases, the steady state's eigenvalue approaches other eigenvalues. For sufficiently large $J'$, the eigenvalues become degenerate and a transition occurs. This intuition also holds for larger systems.



\section{Long chain}\label{sec:long}

Although two atoms already exhibit nonanalytic behavior, in order to discuss phases and phase transitions, we have to consider a long chain of $N$ atoms. We solve this exactly using the Jordan-Wigner transformation, which maps the interacting model to a model of free fermions \cite{lieb61,barouch71,sachdev11}.  Although this is a standard technique for Hermitian systems, there are important differences due to the non-Hermitian nature. Note that the spectra of other non-Hermitian spin chains were considered in Refs.~\cite{refael06,giorgi10,zhang13,hickey13}.

We map each spin to a fermion: $\left|\uparrow\right\rangle=|1\rangle$ and $\left|\downarrow\right\rangle=|0\rangle$. The spin operators are written in terms of fermionic creation and annihilation operators: $\sigma_n^+=c_n^{\dagger}\exp(i\pi\sum_{m<n} c_m^{\dagger}c_m)$ and $\sigma_n^z=2c_n^{\dagger}c_n-1$. After going to Fourier space, $c_n=(e^{-i\pi/4}/\sqrt{N})\sum_ke^{ikn}\tilde{c}_k$, and doing a Bogoliubov transformation, Eq.~\eqref{eq:Heff2} becomes
\begin{eqnarray}
H_{\text{eff}}&=&\sum_k \left[\epsilon(k) \left(\bar{\eta}_k \eta_k - \frac{1}{2}\right)-\frac{i\gamma}{4}\right],\label{eq:Heff_eta}\\
\epsilon(k)&=&\pm\sqrt{(4J\cos k-i\gamma/2)^2 + (4J'\sin k)^2},\label{eq:epsilon}
\end{eqnarray}
in terms of non-Hermitian Bogoliubov quasiparticles,
\begin{eqnarray}
\eta_k&=&u_k \tilde{c}_k + v_k \tilde{c}_{-k}^\dagger, \quad\quad\quad \eta_{-k}=-v_k \tilde{c}_k^\dagger + u_k \tilde{c}_{-k} \nonumber\\
\bar{\eta}_k&=&u_k \tilde{c}_k^\dagger + v_k \tilde{c}_{-k}, \quad\quad\quad \bar{\eta}_{-k}=-v_k \tilde{c}_k + u_k \tilde{c}_{-k}^\dagger. \label{eq:eta}
\end{eqnarray}
The expressions for $u_k$ and $v_k$ are given in Eqs.~\eqref{eq:u_app}--\eqref{eq:v_app}. $\bar{\eta}_k$ and $\eta_k$ obey fermionic statistics: $\{\bar{\eta}_k, \eta_{k'}\}=\delta_{kk'}$ and  $\{\eta_k, \eta_{k'}\}=\{\bar{\eta}_k, \bar{\eta}_{k'}\}=0$. So $\bar{\eta}_k$ and $\eta_k$ act as fermionic creation and annihilation operators. But $\bar{\eta}_k\neq\eta_k^\dagger$ due to the non-Hermitian nature, as seen from Eq.~\eqref{eq:eta}, \eqref{eq:u_app}, and \eqref{eq:v_app}.

Thus, Eq.~\eqref{eq:Heff_eta} is a model of free $\eta$ fermions. We define the vacuum state $|G\rangle$ as the state that satisfies $\eta_k|G\rangle=0$ for all $k$. (But note that $\langle G|\bar{\eta}_k\neq0$.) The eigenstates of $H_\text{eff}$ are given by $|G\rangle$ and its fermionic excitations: $\bar{\eta}_{k_1}\bar{\eta}_{k_2}\cdots\bar{\eta}_{k_n}|G\rangle$.

\begin{figure}[t]
\centering
\includegraphics[width=3.5 in,trim=1.1in 4in 0.9in 4.2in,clip]{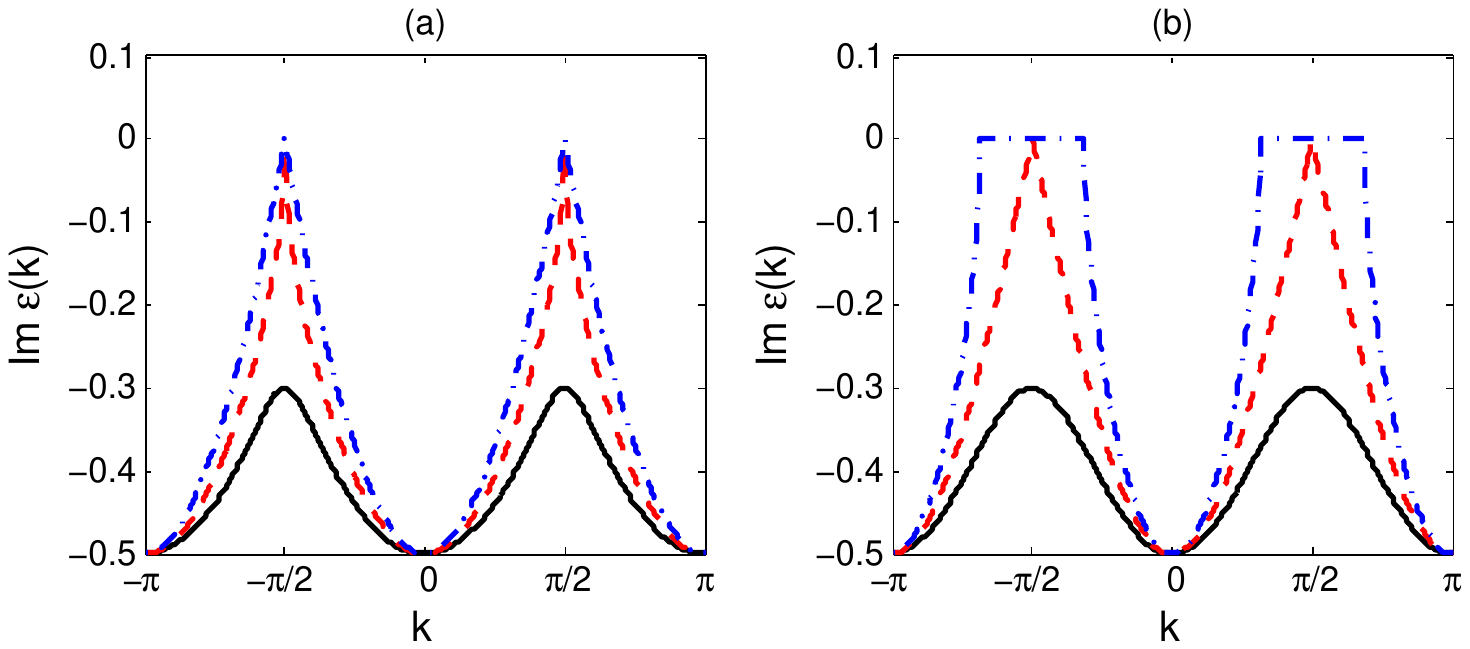}
\caption{\label{fig:spectrum_imag}Imaginary part of $\epsilon(k)$ for $J'=0.1\gamma$ (black solid line), $J'=0.125\gamma$ (red dashed line), and $J'=0.15\gamma$ (blue dash-dotted line). (a) $J=0.1\gamma$. (b) $J=0$.}
\end{figure}

\begin{figure}[t]
\centering
\includegraphics[width=3.5 in,trim=1.1in 4in 0.9in 4.2in,clip]{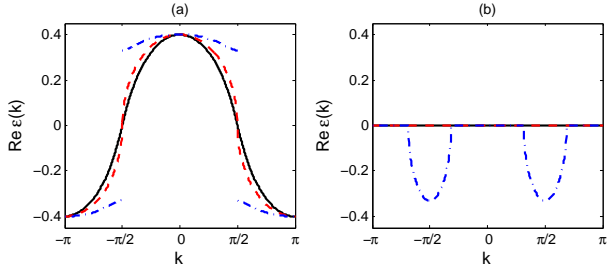}
\caption{\label{fig:spectrum_real}Real part of $\epsilon(k)$ for same parameters as Fig.~\ref{fig:spectrum_imag}.}
\end{figure}

The eigenvalue of $|G\rangle$ is $\lambda_G=-\sum_k[\epsilon(k)+i\gamma/2]/2$. There is freedom in choosing either the $+$ or $-$ branch in Eqs.~\eqref{eq:epsilon} and \eqref{eq:u_app}. In fact, one can choose either branch for each value of $k$. For convenience, we choose the sign convention for each $k$ such that the imaginary part of $\epsilon(k)$ is always negative. This way, $|G\rangle$ is the steady state wavefunction, since all other eigenvalues have smaller (more negative) imaginary parts \footnote{This is solely a matter of convenience. If we chose the sign convention such that the imaginary part of $\epsilon(k)$ is always positive, then the steady state wavefunction would be the eigenstate with fermionic excitations for all $k$. See the discussion after Eq.~(2.18) in Ref.~\cite{lieb61}.}. Figure \ref{fig:spectrum_imag} plots the imaginary part of $\epsilon(k)$, while Fig.~\ref{fig:spectrum_real} plots the real part.

The other eigenvalues are found using Eq.~\ref{eq:epsilon}. For example, the eigenstate $\bar{\eta}_{\pi/2}|G\rangle$ has the eigenvalue $\lambda_G+\epsilon(\pi/2)$. The eigenstate $\bar{\eta}_{\pi/4}\bar{\eta}_{\pi/2}|G\rangle$ has the eigenvalue $\lambda_G+\epsilon(\pi/4)+\epsilon(\pi/2)$. We emphasize that $\epsilon(k)$ does not represent the eigenvalues; rather, it represents the \emph{offset} between eigenvalues.

A phase transition of $|G\rangle$ occurs when $\epsilon(k)=0$ for some $k$, since then an eigenstate is degenerate with $|G\rangle$. As seen from Fig.~\ref{fig:spectrum_imag}, this occurs first for $k=\pm\pi/2$. We define the gap,
\begin{eqnarray}
\Delta&=&\left|\text{Im }\epsilon\left(\frac{\pi}{2}\right)\right|, \label{eq:gap}
\end{eqnarray}
as the difference in imaginary parts between the eigenvalue of $|G\rangle$ and the eigenvalue of $\bar{\eta}_{\pi/2}|G\rangle$ \footnote{Since this gap is independent of $N$, if the gap is nonzero, even a large system will end up in the steady state like in Fig.~\ref{fig:population}.}. The gap indicates how quickly the steady state is reached during the non-Hermitian evolution. The gap closes and a phase transition occurs when $|J'|=\gamma/8$, which is the same as for two atoms.

To characterize the phases, we calculate several observables for $|G\rangle$: $\langle\sigma^z_n\rangle$, $\langle\sigma^x_m\sigma^x_n\rangle$, $\langle\sigma^y_m\sigma^y_n\rangle$, and $\langle\sigma^z_m\sigma^z_n\rangle$. Appendix \ref{sec:app_wick} provides details on how to calculate these observables, which are quite different from the Hermitian case. Note that $\langle\sigma^x_n\rangle=\langle\sigma^y_n\rangle=0$.

We now describe the phases on both sides of the transition. There are two qualitatively different cases, $J\neq0$ and $J=0$, which we consider separately.






\subsection{Case of $J\neq0$}

Figure \ref{fig:spectrum_imag}(a) shows $\text{Im } \epsilon(k)$. When $|J'|<\gamma/8$, $\text{Im }\epsilon(k)$ is always negative, so $|G\rangle$ is gapped, i.e., $\Delta>0$. When $|J'|\geq \gamma/8$, $\text{Im }\epsilon(k)=0$ for $k=\pm\pi/2$, so $|G\rangle$ has gapless excitations.

\begin{figure}[t]
\centering
\includegraphics[width=3.5 in,trim=1.1in 4.1in 0.9in 4.1in,clip]{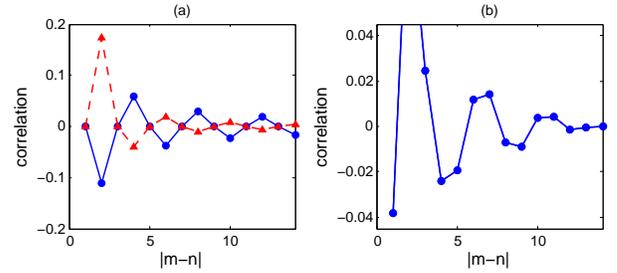}
\caption{\label{fig:corr}Ordered phase's correlation functions: $\langle\sigma^x_m\sigma^x_n\rangle$ (blue solid line) and $\langle\sigma^y_m\sigma^y_n\rangle$ (red dashed line) for $J'=0.13\gamma$ with (a) $J=0.1\gamma$ and (b) $J=0$.}
\end{figure}

Figure \ref{fig:corr}(a) shows the correlation functions for $|G\rangle$. In general, $\langle\sigma^x_m\sigma^x_n\rangle$ and $\langle\sigma^y_m\sigma^y_n\rangle$ are zero for odd distances, which implies that the chain divides into two alternating sublattices with a degree of freedom between them. When $|J'|<\gamma/8$, the correlations $\langle\sigma^x_m\sigma^x_n\rangle$, $\langle\sigma^y_m\sigma^y_n\rangle$, and $\langle\sigma^z_m\sigma^z_n\rangle-\langle\sigma^z_m\rangle\langle\sigma^z_n\rangle$ for even distances decay exponentially with distance. When $|J'|\geq\gamma/8$, they  all decay according to a power law. So within each sublattice, there is short-range order for $|J'|<\gamma/8$ and quasi-long-range order for $|J'|\geq\gamma/8$ [Fig.~\ref{fig:scaling}(a)]. 


The quasi-long-range order here is quite surprising. In Hermitian systems, quasi-long-range order occurs only when the Hamiltonian has a continuous symmetry. However, $H_\text{eff}$ does not have a continuous symmetry when both $J,J'\neq0$.

Figure \ref{fig:corr}(a) shows that the ordered state is frustrated within each sublattice: $\langle\sigma^y_m\sigma^y_n\rangle$ is ferromagnetic with the second neighbor but antiferromagnetic with the fourth neighbor. This is in contrast to $\langle\sigma^x_m\sigma^x_n\rangle$, which shows conventional antiferromagnetic ordering within each sublattice. The frustration of $\langle\sigma^y_m\sigma^y_n\rangle$ is surprising since the lattice is one-dimensional and has only nearest-neighbor interactions. In contrast, Hermitian systems exhibit frustration when the lattice is triangular \cite{ramirez94} or the interaction is long-range \cite{selke88}.

As $|J'|$ increases towards $\gamma/8$, the correlation length $\xi$ diverges as $\sim(\gamma/8-|J'|)^{-\nu}$ where $\nu$ is the critical exponent (Fig.~\ref{fig:corr_length}). In Appendix \ref{sec:app_corr}, we analytically calculate $\langle\sigma^z_m\sigma^z_n\rangle-\langle\sigma^z_m\rangle\langle\sigma^z_n\rangle$ for large distances and show that $\nu=1$, which is the same as for Hermitian spin models in one dimension \cite{sachdev11}. Numerically, we find that $\langle\sigma^x_m\sigma^x_n\rangle$ and $\langle\sigma^y_m\sigma^y_n\rangle$ also have $\nu=1$.

The dynamical critical exponent $z$ is given by the divergence of the gap at the phase transition \cite{sachdev11}. We find from Eqs.~\eqref{eq:epsilon} and \eqref{eq:gap} that the gap diverges as $\Delta\sim(\gamma/8-|J'|)^{1/2}$. Thus, $z\nu=1/2$ and $z=1/2$.

\subsection{Case of $J=0$}

Figure \ref{fig:spectrum_imag}(b) shows $\text{Im }\epsilon(k)$ for this case. When $|J'|<\gamma/8$, the steady state is gapped as before. When $|J'|=\gamma/8$, the gap closes at $k=\pm\pi/2$. But when $|J'|>\gamma/8$, $\text{Im }\epsilon(k)=0$ for an extended range around $k=\pm\pi/2$, which is an important difference with the $J\neq 0$ case.

Figure \ref{fig:corr}(b) shows the correlation functions for $|G\rangle$. In general, $\langle\sigma^x_m\sigma^x_n\rangle=\langle\sigma^y_m\sigma^y_n\rangle$. When $|J'|<\gamma/8$, $\langle\sigma^x_m\sigma^x_n\rangle$, $\langle\sigma^y_m\sigma^y_n\rangle$, and $\langle\sigma^z_m\sigma^z_n\rangle-\langle\sigma^z_m\rangle\langle\sigma^z_n\rangle$ are zero for odd distances, and their values for even distances decay exponentially in distance. But when $|J'|\geq\gamma/8$, they are nonzero for all distances and form a spin-density-wave pattern, whose magnitude decays according to a power law [Fig.~\ref{fig:scaling}(b)].

In Appendix \ref{sec:app_corr}, we show analytically that the correlation length $\xi$ diverges with critical exponent $\nu=1/2$. It is surprising that this is not 1, which is the value for $J\neq0$ as well as for Hermitian spin chains. From the divergence of Eq.~\eqref{eq:gap}, we again find $z\nu=1/2$, so $z=1$.

Figure \ref{fig:sz}(b) plots $\langle\sigma^z_n\rangle$. When $|J'|\leq\gamma/8$, $\langle\sigma^z_n\rangle=-\frac{2}{\pi}E(64J'^2)$, where $E(x)$ is the complete elliptic integral of the second kind. Interestingly, $d\langle\sigma^z_n\rangle/dJ'$ exhibits a logarithmic divergence at $|J'|=\gamma/8$. (This singularity does not occur when $J\neq0$.)

\begin{figure}[t]
\centering
\includegraphics[width=3.5 in,trim=1.1in 4.1in 0.9in 4.1in,clip]{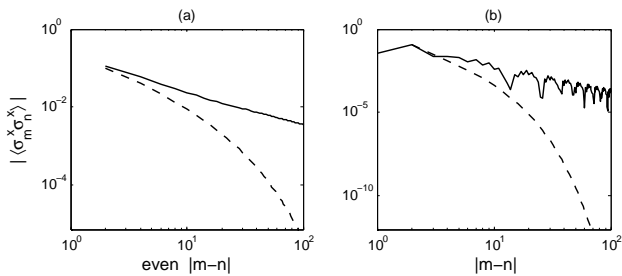}
\caption{\label{fig:scaling}Correlation functions for $J'=0.13\gamma$ (solid line) and $J'=0.12\gamma$ (dashed line) for (a) $J=0.1\gamma$ and (b) $J=0$. Panel (b) shows only even distances for $J'=0.12\gamma$.}
\end{figure}

\begin{figure}[t]
\centering
\includegraphics[width=3. in,trim=2.2in 4.1in 2in 4.2in,clip]{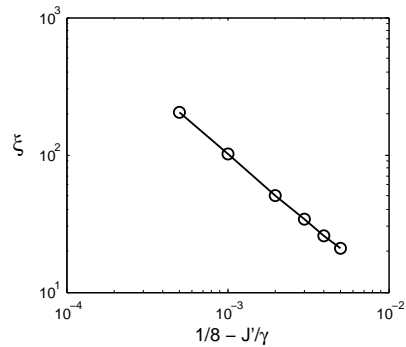}
\caption{\label{fig:corr_length}Correlation length $\xi(J')$ for $J=0.1\gamma$, found by fitting exponential decay of $\langle\sigma^x_m\sigma^x_n\rangle$. At the phase transition, $\xi$ diverges with critical exponent $\nu=1$.}
\end{figure}

\subsection{Exact diagonalization for a small chain} \label{sec:ed}

\begin{figure}[t]
\centering
\includegraphics[width=3.5 in,trim=1.1in 4.1in 0.9in 4.1in,clip]{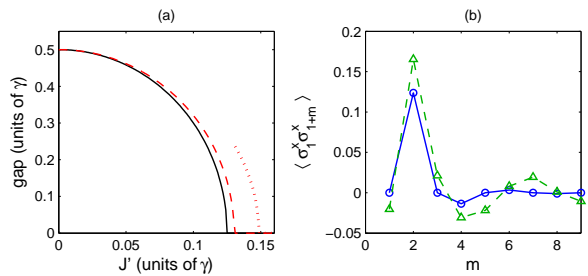}
\caption{\label{fig:ed}Exact diagonalization of $N=10$ atoms with open boundary conditions and $J=0$. (a) Gap (red dashed line) and second gap (red dotted line). Gap for $N=\infty$ (solid black line) from Eq.~\eqref{eq:gap}. (b) Correlations between first atom and other atoms for $J'=0.12\gamma$ (blue solid line, circles) and $J'=0.14\gamma$ (green dashed line, triangles).}
\end{figure}

The above results assumed an infinite chain. In this section, we consider a small chain of $N=10$ with open boundary conditions, since it is the more experimentally relevant situation. Figure \ref{fig:ed}(a) shows the gap, defined as the difference in imaginary parts between the eigenvalue with the largest imaginary part and the eigenvalue with the second largest imaginary part. In the limit of $N\rightarrow\infty$, we should recover Eq.~\eqref{eq:gap}. Figure \ref{fig:ed}(a) shows that the gap for $N=10$ is very similar to that for $N=\infty$. It is important to notice that already for $N=10$, the gap closes, leading to non-analytic change of the steady state; when $J=0$, the gap closes at $J'= 0.131\gamma$. (But recall from Sec.~\ref{sec:N2} that the gap closes even for $N=2$.)

Before the gap closes, there is a unique steady state. After the gap closes, there are four eigenvalues with the largest imaginary part, meaning that there are four steady states. The eigenvalues have different real parts, which allows them to be differentiated via Fourier analysis, as discussed in Sec.~\ref{sec:experiment}. We now define the ``second gap'' as the difference in imaginary parts between the four eigenvalues with the largest imaginary part and the eigenvalue with the fifth largest imaginary part. The second gap indicates how quickly the system settles into the manifold of four steady states. Figure \ref{fig:ed}(a) shows that the second gap is nonzero for an interval of $J'$. The existence of the second gap makes it experimentally easier to distinguish between the steady states, since there are only four of them.

Figure \ref{fig:ed}(b) shows the correlations between the first atom and the other atoms. There is a visible difference in the correlations before and after the phase transition. Before the transition, the correlations are zero for odd distances, while after the transition, the correlations are always nonzero due to the emergence of the spin-density wave. One can also see that the correlation decays slower after the phase transition, just like in an infinite chain.

In summary, with $N=10$ atoms, the behavior already resembles that of an infinite system. In particular, there is a transition from short-range order to the spin-density wave. This is the experimental advantage of non-Hermitian models: there are sharp transitions even for small systems, so one does not need a large number of atoms as in Hermitian models.

\section{Comparison with Hermitian model and master equation}

The ground state of the Hermitian model [Eq.~\eqref{eq:H}] is either ferromagnetic or antiferromagnetic \cite{lieb61}. There is long-range order, except when $|J_x|=|J_y|$, in which case there is quasi-long-range order. If $H$ included a real field $\sum_n(h/4)\sigma_n^z$, then there would be short-range order when $|J|<h/8$ with $\nu=1$ \cite{barouch71}.

The steady-state density matrix of the master equation [Eq.~\eqref{eq:master}] has only short-range order in one dimension \cite{joshi13}. Mean-field theory predicts that when $J_x J_y>-\gamma^2/64$, the system is in the staggered-$XY$ phase, characterized by $\langle\sigma^x_m\sigma^x_n\rangle$ and $\langle\sigma^y_m\sigma^y_n\rangle$ being 0 for odd distances and both positive for even distances \cite{lee13c}.

The magnetic behavior of the non-Hermitian model is qualitatively different from both the Hermitian model and the master equation. It has a phase transition in one dimension and in fact already exhibits a sharp transition for two atoms. There is quasi-long-range order for $|J'|>\gamma/8$, which is an extended area in $J_x,J_y$ space instead of a line as in the Hermitian case. $\nu$ can be 1/2, which is different from the Hermitian case. The ordered phases are also different. Since $\langle\sigma^x_m\sigma^x_n\rangle$ and $\langle\sigma^y_m\sigma^y_n\rangle$ always have opposite signs when $J\neq0$, the ordered phase is different from both the staggered-$XY$ phase and the antiferromagnetic phase. In addition, the spin-density-wave phase (for $J=0$) is not present in either the Hermitian model \cite{lieb61} or the master equation \cite{lee13c,joshi13}.

We emphasize that $H_\text{eff}$ has different critical behavior than the master equation even though both include dissipation. Recent works indicate that phase transitions of master equations are in the universality class of classical phase transitions \cite{sieberer13}, and thus should not have phase transitions in one dimension. Since $H_\text{eff}$ does have a phase transition in one dimension, it is a \emph{quantum} phase transition \cite{sachdev11}. 



\section{Experimental implementation} \label{sec:experiment}

The Hermitian $XY$-model [Eq.~\eqref{eq:H}] can be implemented using trapped ions \cite{molmer99,porras04,britton12,richerme14}, an array of cavities \cite{bardyn12,joshi13}, atoms within a cavity \cite{morrison08}, and Rydberg atoms in an optical lattice \cite{bouchoule02,lee13c,viteau11,li13}. These experimental setups are able to tune $J$ and $J'$. To implement the non-Hermitian model [Eq.~\eqref{eq:Heff}], one would optically pump $\left|\uparrow\right\rangle$ into the auxiliary state $|a\rangle$. 

We discuss a specific realization using trapped ${}^{171}\text{Yb}^+$ ions [Fig.~\ref{fig:level_scheme}(c)]. Let $\left|\downarrow\right\rangle$ be ${|S_{1/2},F=0\rangle}$, let $\left|\uparrow\right\rangle$ be ${|D_{3/2},F=2\rangle}$, and let $|a\rangle$ be ${|S_{1/2},F=1\rangle}$. One would optically pump $\left|\uparrow\right\rangle$ to ${|P_{3/2},F=2\rangle}$, which decays into $|a\rangle$ instead of $\left|\downarrow\right\rangle$ due to dipole selection rules. (The excitation to ${|P_{3/2},F=2\rangle}$ should be much weaker than the decay so that the level can be adiabatically eliminated.) $J$ and $J'$ can be on the order of $2\pi\times 1 \text{ kHz}$ \cite{richerme14}, and one can set $\gamma=2\pi\times 10 \text{ kHz}$ so that $J,J'\sim 0.1\gamma$. To detect whether an atom has decayed to $|a\rangle$, one would laser-excite $|a\rangle$ to $|P_{1/2},F=0\rangle$ and observe the fluorescence; the absence of fluorescence means the atom has not decayed. If one finds that no atom has decayed, one stops the laser-excitation of $|a\rangle$ and then proceeds to measure correlation functions.

To observe the phase transition, one would measure the correlations of the steady state for $|J'|<\gamma/8$ and $|J'|>\gamma/8$ to see the onset of quasi-long-range order. If the gap [Eq.~\eqref{eq:gap}] is so small that reaching the steady state is difficult, one can  use Fourier analysis to distinguish between the eigenstates. Each eigenstate of $H_\text{eff}$ oscillates with a frequency given by the real part of its eigenvalue. Figure \ref{fig:spectrum_real} shows that $|G\rangle$ and other long-lived eigenstates have different oscillation frequencies. Thus, when the wavefunction is a superposition of these eigenstates, observables will be time-dependent due to interference of the different frequencies. By measuring the observables over time and then taking the Fourier transform, one can identify each long-lived eigenstate.


We now estimate the probability that a given experimental run has no decay to $|a\rangle$. 
Each run should be long enough so that the system has converged into the steady state. The timescale is estimated by $1/\Delta$, where $\Delta$ is the gap [Eq.~\eqref{eq:gap}]. The number of decay events during this time interval is Poissonian with an average of $\mu\approx\gamma N(\langle \sigma_n^z\rangle+1)/2\Delta$, so the probability of no decay events during a run is $P\approx e^{-\mu}$. For example, suppose $J'=J=0.1\gamma$ and $\gamma=2\pi\times 10 \text{ kHz}$. Then each run lasts for about 50 $\mu$s. For $N=2$, $P\approx 0.3$. For $N=20$, $P\approx 0.007$. This success probability is several orders of magnitude higher than that for heralded entanglement protocols \cite{chou05,matsukevich06,moehring07,casabone13}, and $N=20$ atoms is enough to see the results described in this paper as discussed in Sec.~\ref{sec:ed}. We emphasize that sharp transitions occur for any $N\geq2$.


\section{Conclusion}

The non-Hermitian model exhibits phases and phase transitions that are absent from the Hermitian model and master equation. Thus, non-Hermitian quantum mechanics is a promising route to find new condensed-matter phenomena. For future work, perhaps one can map a $D$-dimensional non-Hermitian quantum model to a $(D+1)$-dimensional classical model like in the Hermitian case \cite{sachdev11}. One can also see whether the entanglement between atoms exhibits critical behavior at the phase transition, as in the Hermitan model \cite{osterloh02} and master equation \cite{joshi13}. Furthermore, one can consider the effect of disorder using real-space renormalization group \cite{refael06}. Finally, it would be interesting to study the Lee-Yang zeros of the non-Hermitian model \cite{wei12,heyl13,wei14,peng14,hickey13}.


\section{Acknowledgements}

We thank S. Sachdev, G. Refael, N. Moiseyev, S. Gopalakrishnan, and M. Foss-Feig for useful discussions. This work was supported by the NSF through a grant to ITAMP. C.K.C.~is supported by the Croucher Foundation.

\appendix
\onecolumngrid

\section{Origin of the non-Hermitian Hamiltonian} \label{sec:app_nonh}

Here, we explain why the wavefunction evolves with a non-Hermitian Hamiltonian in the absence of a decay event. This was thoroughly discussed in previous works \cite{dalibard92,molmer93,dum92,wiseman96,plenio98}, but we sketch out the derivation here for the reader's convenience. For simplicity, we consider only one atom. But in addition to the atomic wave function, we keep track of the number of photons around the atom. The photonic state is $|n\rangle$, where $n$ is the number of photons.

Suppose the atom starts in a superposition,
\begin{eqnarray}
|\psi(t)\rangle&=&(\alpha\left|\downarrow\right\rangle+\beta\left|\uparrow\right\rangle)|0\rangle.\label{eq:psiphot}
\end{eqnarray}
In a short time interval $dt$, the probability that the atom decays is $p=\gamma|\beta|^2dt \ll 1$. The wave function then evolves to
\begin{eqnarray}
|\psi(t+dt)\rangle&=&\alpha\left|\downarrow\right\rangle|0\rangle+\beta\left(1-\frac{\gamma\, dt}{2}\right)\left|\uparrow\right\rangle|0\rangle+\sqrt{p}|a\rangle|1\rangle.\label{eq:psiphotdt}
\end{eqnarray}
In other words, $\left|\uparrow\right\rangle$ decays with probability $p$ to $|a\rangle$, emitting a photon in the process. 

At this point, the environment detects whether or not there is a photon. If it detects a photon, the $|1\rangle$ component of Eq.~\eqref{eq:psiphotdt} is projected out:
\begin{eqnarray}
|\psi(t+dt)\rangle&=&|a\rangle|1\rangle. \label{eq:psi_a1}
\end{eqnarray}
If no photon is detected, the $|0\rangle$ component is projected out (and normalized):
\begin{eqnarray}
|\psi(t+dt)\rangle&=&\alpha\left(1+\frac{\gamma|\beta|^2dt}{2}\right)\left|\downarrow\right\rangle|0\rangle+\beta\left(1-\frac{\gamma|\alpha|^2 dt}{2}\right)\left|\uparrow\right\rangle|0\rangle.\label{eq:psiphotdt0}
\end{eqnarray}
Comparing Eqs.~\eqref{eq:psiphot} and \eqref{eq:psiphotdt0}, we see that the population in $\left|\uparrow\right\rangle$ decreased a little, while the population in $\left|\downarrow\right\rangle$ increased a little. In other words, \emph{the non-detection of a photon shifts the atom towards $\left|\downarrow\right\rangle$ in a nonunitary way}. To account for this effect, we add the non-Hermitian term to the Hamiltonian [Eq.~\eqref{eq:Heff}]. Thus, in the absence of a decay event, the atom evolves with $H_{\text{eff}}$.

Now, how can the experimentalist measure whether a decay happened? Detecting the single photon in Eq.~\eqref{eq:psi_a1} is difficult. A much better way is to measure the population in $|a\rangle$ by laser-exciting it to another state [Fig.~\ref{fig:level_scheme}(d)]. If the atom is in $|a\rangle$, it will scatter \emph{many} photons, allowing one to measure the population with almost 100\% efficiency \cite{myerson08}.

Note that it is easier to measure $|a\rangle$ after the experimental run than during. Before doing the measurement, one would set $\gamma=0$ by turning off the optical pumping laser. This way, a decay will not occur during the measurement. (It takes some time to achieve a high measurement efficiency \cite{myerson08}.)

\section{Jordan-Wigner transformation} \label{sec:app_jw}

Here, we provide more details of the Jordan-Wigner calculation for the non-Hermitian model, since there are some important differences with the Hermitian model.
After doing the Jordan-Wigner transformation and going to Fourier space, $c_n=(e^{-i\pi/4}/\sqrt{N})\sum_ke^{ikn}\tilde{c}_k$, Eq.~(4) of the main text becomes
\begin{eqnarray}
H_{\text{eff}}&=&\sum_{k>0} \Bigg[
\left( \begin{array}{cc} \tilde{c}_k^{\dagger} & \tilde{c}_{-k} \end{array} \right)
M_k
\left( \begin{array}{c} \tilde{c}_k \\ \tilde{c}_{-k}^{\dagger} \end{array} \right)
-\frac{i\gamma}{2}\Bigg], \\
M_k&=&\left( \begin{array}{cc} 4J\cos k-i\gamma/2 & -4J'\sin k \\ -4J'\sin k & -(4J\cos k-i\gamma/2) \end{array} \right).
\end{eqnarray}
Let the right eigenvectors of $M_k$ be $(u_k,v_k)^T$ and $(-v_k,u_k)^T$, where
\begin{eqnarray}
u_k&=&\frac{i\gamma-8J\cos k \pm 2\sqrt{(4J\cos k - i\gamma/2)^2+(4J'\sin k)^2}}{\mathcal{C}}, \label{eq:u_app}\\
v_k&=&\frac{8J'\sin k}{\mathcal{C}},\label{eq:v_app}
\end{eqnarray}
where the normalization constant $\mathcal{C}$ is such that $u_k^2+v_k^2=1$. The sign convention in Eq.~\eqref{eq:u_app} is the same as for $\epsilon(k)$, i.e., the sign convention for each $k$ is such that the imaginary part of $\epsilon(k)$ is negative.

Then after diagonalizing $M_k$, we obtain
\begin{eqnarray}
H_{\text{eff}}&=&\sum_k \left[\epsilon(k) \left(\bar{\eta}_k \eta_k - \frac{1}{2}\right)-\frac{i\gamma}{4}\right],\label{eq:Heff_eta_app}\\
\epsilon(k)&=&\pm\sqrt{(4J\cos k-i\gamma/2)^2 + (4J'\sin k)^2},\label{eq:epsilon_app}
\end{eqnarray}
in terms of non-Hermitian Bogoliubov quasiparticles,
\begin{eqnarray}
\eta_k&=&u_k \tilde{c}_k + v_k \tilde{c}_{-k}^\dagger, \quad\quad\quad \eta_{-k}=-v_k \tilde{c}_k^\dagger + u_k \tilde{c}_{-k}, \nonumber\\
\bar{\eta}_k&=&u_k \tilde{c}_k^\dagger + v_k \tilde{c}_{-k}, \quad\quad\quad \bar{\eta}_{-k}=-v_k \tilde{c}_k + u_k \tilde{c}_{-k}^\dagger. \label{eq:eta_app}
\end{eqnarray}



The vacuum state $|G\rangle$ is defined via $\eta_k|G\rangle=0$, and is given explicitly by
\begin{eqnarray}
|G\rangle&=&\frac{1}{\sqrt{\mathcal{N}}}\prod_{k>0} (u_k - v_k c_k^\dagger c_{-k}^\dagger) |0\rangle, \label{eq:G_app}
\end{eqnarray}
where $\mathcal{N}=\prod_{k>0}(|u_k|^2+|v_k|^2)$ is the normalization constant. Equation \eqref{eq:G_app} is similar to the BCS ground state (see Pg.~272 of Ref.~\cite{altland10}). It is easy to check that $\eta_k|G\rangle=0$. Note that $|u_k|^2+|v_k|^2\neq 1$ for the non-Hermitian case. From Eq.~\eqref{eq:G_app}, one finds
\begin{eqnarray}
\langle\sigma^z_n\rangle &=& \frac{1}{\pi}\int_0^\pi {dk \frac{-|u_k|^2+|v_k|^2}{|u_k|^2+|v_k|^2}}.\label{eq:sz_app}
\end{eqnarray}

\section{Wick expansion} \label{sec:app_wick}

Now we want to calculate
\begin{eqnarray}
\langle\sigma^x_m\sigma^x_n\rangle&=&\langle B_m A_{m+1} B_{m+1} \cdots A_{n-1} B_{n-1} A_n\rangle, \label{eq:sxsx_app}\\
\langle\sigma^y_m\sigma^y_n\rangle&=&(-1)^{n-m}\langle A_m B_{m+1} A_{m+1} \cdots B_{n-1} A_{n-1} B_n\rangle,\quad \label{eq:sysy_app}\\
\langle\sigma^z_m\sigma^z_n\rangle&=&\langle A_m B_m A_n B_n\rangle,\label{eq:szsz_app}
\end{eqnarray}
for $|G\rangle$, where $A_n=c_n^\dagger+c_n$ and $B_n=c_n^\dagger-c_n$. To calculate these, we use Wick's theorem. However, when applying Wick's theorem, it is convenient to use $\eta_k,\eta_k^\dagger$ instead of $\eta_k,\bar{\eta}_k$, since $\langle G|\eta_k^\dagger=0$ while $\langle G|\bar{\eta}_k\neq0$. So we note that
\begin{eqnarray}
\tilde{c}_k=\frac{u_k^*\eta_k - v_k\eta_{-k}^\dagger}{|u_k|^2+|v_k|^2}, \quad\quad\quad \tilde{c}_{-k}=\frac{u_k^*\eta_{-k} + v_k\eta_k^\dagger}{|u_k|^2+|v_k|^2},
\end{eqnarray}
and $\langle\eta_k\eta_k^\dagger\rangle=|u_k|^2+|v_k|^2$.
Then we do a Wick expansion of Eqs.~\eqref{eq:sxsx_app}--\eqref{eq:szsz_app} in terms of contractions of operator pairs. For example,
\begin{eqnarray}
\langle\sigma^z_m\sigma^z_n\rangle&=&\langle A_m B_m\rangle\langle A_n B_n\rangle - \langle A_m B_n\rangle\langle A_n B_m\rangle - \langle A_m A_n\rangle\langle B_m B_n\rangle\\
&=&\langle\sigma^z_m\rangle\langle\sigma^z_n\rangle - \langle A_m B_n\rangle\langle A_n B_m\rangle - \langle A_m A_n\rangle\langle B_m B_n\rangle.
\end{eqnarray}

The pair contractions for $A_n$ and $B_n$ are
\begin{eqnarray}
\langle A_m A_n\rangle&=&\delta_{mn}+\frac{1}{\pi}\int_0^\pi dk \sin k(n-m) \left(\frac{u_k v_k^* - u_k^* v_k}{|u_k|^2+|v_k|^2}\right),\\
\langle B_m B_n\rangle&=&-\delta_{mn}+\frac{1}{\pi}\int_0^\pi dk \sin k(n-m) \left(\frac{u_k v_k^* - u_k^* v_k}{|u_k|^2+|v_k|^2}\right),\\
\langle B_m A_n\rangle&=&-\langle A_n B_m\rangle \\
&=&-\frac{1}{\pi}\int_0^\pi dk \cos k(n-m)\left(\frac{|u_k|^2-|v_k|^2}{|u_k|^2+|v_k|^2}\right)+\frac{1}{\pi}\int_0^\pi dk \sin k(n-m)\left(\frac{u_k v_k^* + u_k^* v_k}{|u_k|^2+|v_k|^2}\right) .
\end{eqnarray}
These expressions are different from the Hermitian case \cite{lieb61}. In particular, $\langle A_m A_n\rangle$ and $\langle B_m B_n\rangle$ can be nonzero even when $m\neq n$. (This prevents us from writing $\langle\sigma^x_m\sigma^x_n\rangle$ in terms of a Toeplitz determinant and using Szeg\"{o}'s theorem \cite{barouch71,sachdev11}.)

Since $\langle\sigma^x_m\sigma^x_n\rangle$ and $\langle\sigma^y_m\sigma^y_n\rangle$ may contain many operators, it is useful to write them in terms of the Pfaffian of a skew-symmetric matrix \cite{barouch71}:
\begin{eqnarray}
\langle\sigma^x_m\sigma^x_n\rangle&=&\text{pf}\left(
\begin{array}{ccccccccc}
0\quad\quad & \langle B_m B_{m+1}\rangle & \langle B_m B_{m+2}\rangle     & \cdots & \langle B_m B_{n-1}\rangle     & \langle B_m A_{m+1}\rangle     & \langle B_m A_{m+2}\rangle     & \cdots & \langle B_m A_n\rangle     \\
  & 0           & \langle B_{m+1} B_{m+2}\rangle & \cdots & \langle B_{m+1} B_{n-1}\rangle & \langle B_{m+1} A_{m+1}\rangle & \langle B_{m+1} A_{m+2}\rangle & \cdots & \langle B_{m+1} A_n\rangle \\
  &             & 0               & \cdots & \vdots          & \vdots          & \vdots          & \vdots & \vdots      \\
  &             &                 &        & \langle B_{n-2} B_{n-1}\rangle & \langle B_{n-2} A_{m+1}\rangle & \langle B_{n-2} A_{m+2}\rangle & \cdots & \langle B_{n-2} A_n\rangle \\
  &             &                 &        & 0               & \langle B_{n-1} A_{m+1}\rangle & \langle B_{n-1} A_{m+2}\rangle & \cdots & \langle B_{n-1} A_n\rangle \\
  &             &                 &        &                 & 0               & \langle A_{m+1} A_{m+2}\rangle & \cdots & \langle A_{m+1} A_n\rangle \\
  &             &                 &        &                 &                 &  0              & \cdots & \vdots      \\
  &             &                 &        &                 &                 &                 &        & \langle A_{n-1} A_n\rangle \\
  &             &                 &        &                 &                 &                 &        & 0           
\end{array}
\right)\nonumber,\\
\end{eqnarray}
and similarly for $\langle\sigma^y_m\sigma^y_n\rangle$. The elements on the bottom left are given by the skew-symmetry. This matrix has dimensions $2|m-n|\times2|m-n|$. The Pfaffian of a matrix can be efficiently computed using the fact that $\text{pf}(D)^2=\text{det}(D)$, but this method does not give the sign of the Pfaffian. If one needs the sign of $\langle\sigma^x_m\sigma^x_n\rangle$, it is necessary to explicitly calculate the Wick expansion, which is computationally a lot slower.

\section{Asymptotic behaviors of the correlation functions} \label{sec:app_corr}

In this section, we analytically evaluate the asymptotic behaviors of the $z$-component correlation function,
\begin{eqnarray}
C^{zz}(x)&=&\langle\sigma^z_0 \sigma^z_x \rangle -\langle \sigma^z_0 \rangle \langle \sigma^z_x \rangle \nonumber\\
&=& -\langle A_0 B_x\rangle \langle A_x B_0 \rangle -\langle A_0 A_x\rangle \langle B_0 B_x \rangle,
\end{eqnarray}
for long distances. The main goal is to calculate the critical exponent $\nu$, which describes the divergence of the correlation length as $|J'|$ increases toward $\gamma/8$. We consider the cases of $J=0$ and $J\neq0$ separately.

\subsection{Case of $J=0$}

\begin{figure}[t]
\centering
\includegraphics[width=.6\columnwidth]{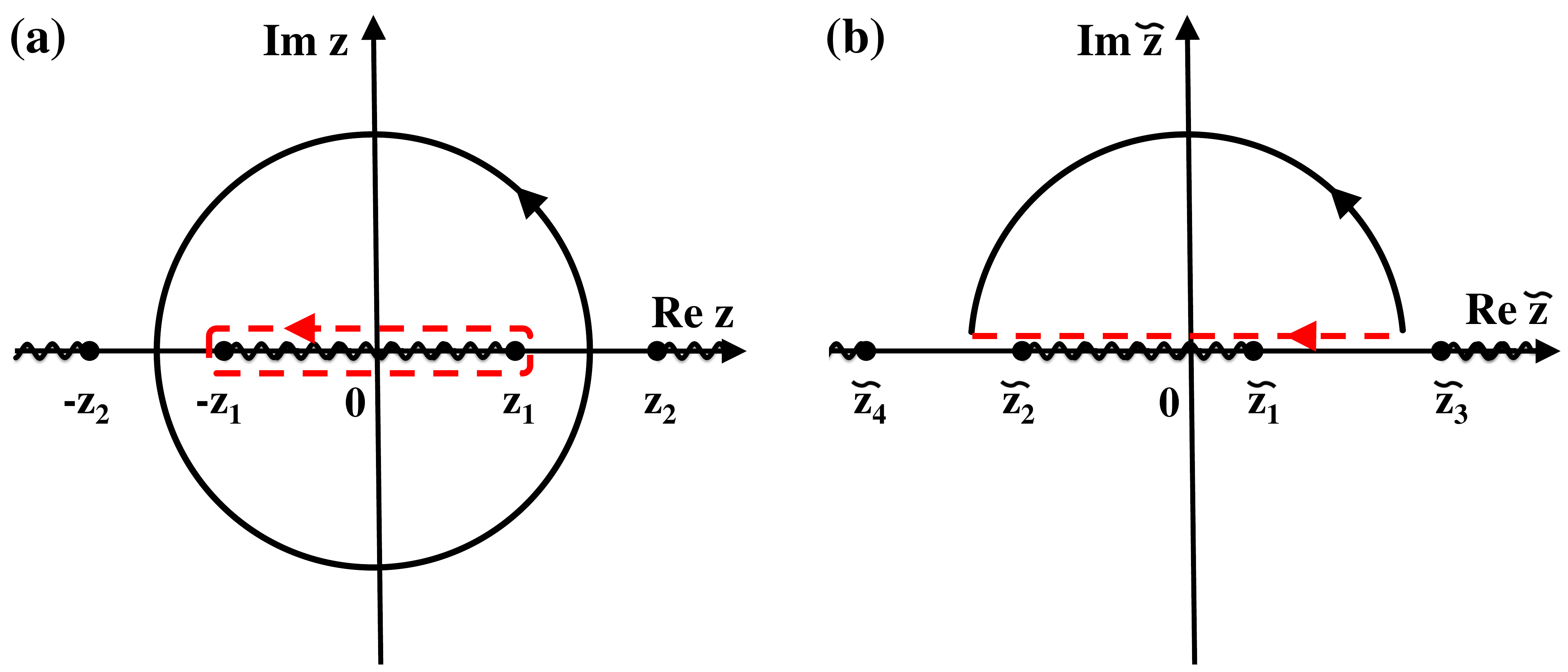}
\caption{Contours for the evaluation of pair contraction functions for (a) $J=0$ and (b) an example of $J\neq 0$. Solid lines are the original contours. Dashed lines are the deformed contours. Jagged lines are the branch cuts.}
\label{fig_branchcut}
\end{figure}

In this case, the sign conventions of $\epsilon_k$ and $u_k$  are independent of $k$. When $|x|>1$, we have $\langle A_0 A_x \rangle = \langle B_0 B_x \rangle =0 $  and $C^{zz}(x)= - \langle B_0 A_x \rangle^2$, where
\begin{eqnarray}
\langle B_0 A_x \rangle=-\frac{1}{\pi}\int^{\pi}_0 dk \cos kx \sqrt {1- \left(\frac{8J'}{\gamma}\right)^2 \sin^2 k},
\label{eq_BA1}
\end{eqnarray}
which is non-zero only for even $x$. We follow the procedure introduced in Ref.~\cite{wu61} to evaluate the above integral in the limit $|x| \gg 1$. We first transform Eq.~(\ref{eq_BA1}) into a contour integral by a change of variable $z=i e^{ik}$, so that
\begin{eqnarray}
\label{eq_BA2}
\langle B_0 A_x \rangle&=&\frac{2 i^x |J'|}{\pi \gamma}\oint dz z^{x-1} \sqrt {\frac{(z^2-z_1^2)(z^2-z_2^2)}{z^2}},  \\
z_1 &=& \frac{\gamma}{J'}\left[\frac{1}{8}-\sqrt{\left(\frac{1}{8}\right)^2 - \left(\frac{J'}{\gamma}\right)^2}\right] , \nonumber \\
z_2 &=& \frac{\gamma}{J'}\left[\frac{1}{8}+\sqrt{\left(\frac{1}{8}\right)^2 - \left(\frac{J'}{\gamma}\right)^2}\right] , \nonumber
\end{eqnarray}
where the contour is along the unit circle [solid line in Fig.~\ref{fig_branchcut}(a)] and the integral is real. The branch points are along the real axis because we have defined $z=i e^{ik}$. The branch cuts are along several segments of the real axis [jagged lines in Fig.~\ref{fig_branchcut}(a)]. The branch cuts are chosen so that the integrand is continuous along the contour. We then deform the contour to lines between $-z_1$ and $z_1$ [dashed line in Fig.~\ref{fig_branchcut}(a)]. Making a further change of variable $y=z/z_1$, we obtain the integral expression,
\begin{eqnarray}
\langle B_0 A_x \rangle=-\frac{8 i |J'|}{\pi \gamma} (i z_1)^{x+1}\int^1_0 dy y^{x-2} \sqrt {(y^2-1)\left[y^2-\left(\frac{z_2}{z_1}\right)^2\right] }.
\label{eq_BA3}
\end{eqnarray}

Now, we expand the above integral for large $x$. As in Ref.~\cite{wu61}, due to the presence of the $y^{x-2}$ term, the integrand is dominated by when $y$ is close to $1$. Expanding around $y=1$ results in a $1/x$ expansion of the overall integral:
\begin{eqnarray}
\langle B_0 A_x \rangle &\doteq& -\frac{8 i |J'|}{\pi \gamma} (i z_1)^{x+1} \sum_{n=0} \frac{(-1)^{n+1}(2n-3)!!}{2^n n!} \left[ \left(\frac{z_2}{z_1}\right)^2 -1 \right]^{\frac{1-2n}{2}} \times \int^1_0 dy y^{x-2}(1-y^2)^{\frac{2n+1}{2}} \nonumber \\
&=&  -\frac{4 i |J'|}{\pi \gamma} (i z_1)^{x+1} \sum_{n=0} \frac{(-1)^{n+1}(2n-3)!!}{2^n n!} \left[ \left(\frac{z_2}{z_1}\right)^2 -1 \right]^{\frac{1-2n}{2}} \times  \frac{\Gamma(n+\frac{3}{2})\Gamma(\frac{x-1}{2})}{ \Gamma(\frac{x}{2}+n+1)} \nonumber\\
&=& -4 i\sqrt \frac{ 2 }{ \pi}\frac{|J'|}{\gamma} (i z_1)^{x+1}x^{-\frac{3}{2}}\left [\left(\frac{z_2}{z_1}\right)^2-1\right]^{\frac{1}{2}} \left \{1+\frac{3}{2 x} \left [\left(\frac{z_2}{z_1}\right)^2-1\right]^{-1} +\mathcal O \left(\frac{1}{x^2}\right)\right\},
\label{eq_BA4}
\end{eqnarray}
where $\doteq$ means that the two expressions are asymptotically the same. Consequently, the $z$-component correlation has the following asymptotic behavior for even $x$:
\begin{eqnarray}
C^{zz}(x\gg 1) \doteq -\frac{32 }{\pi} \left(\frac{J'}{\gamma}\right)^2 x^{-3}z_1^{2x+2} \times \left[ \left(\frac{z_2}{z_1}\right)^2-1 +\frac{3}{x} + \mathcal O \left(\frac{1}{x^2}\right)\right],
\label{eq_BA5}
\end{eqnarray}
or $C^{zz}(x) \sim x^{-3} e^{-x/\xi}$ with $\xi = -1/(2\ln z_1)$. Near the critical point,
\begin{eqnarray}
\xi = \frac{1}{8 (\frac{1}{8}-\frac{|J'|}{\gamma})^{\frac{1}{2}}},
\end{eqnarray}
so the critical exponent $\nu$ is 1/2.

\subsection{Case of $J \neq 0$}

The asymptotic behavior for the $J \neq 0 $ case can be carried out in a similar manner. We shall consider only even $x$. For $|x| > 1 $, we find:
\begin{eqnarray}
\langle B_0 A_{\pm x} \rangle =\text{Re}\left\{-\frac{\text{sgn}(J)}{\pi}\int^{\frac{\pi}{2}}_{-\frac{\pi}{2}} dk  \left[-i \cos kx \pm \sin kx \left(\frac{-1+8 i J \cos k/\gamma}{8 J' \sin k/\gamma}\right)\right] \sqrt{(i-8J \cos k/\gamma)^2 +(8J' \sin k/\gamma)^2}\right\},\nonumber \\
\label{eq_BA1_nonzeroJ}
\end{eqnarray}
where $\text{sgn}(J)$ is the sign of $J$, $\langle A_0 A_x \rangle = \langle B_0 B_x \rangle=0$, and $\langle A_0 B_x \rangle = -\langle B_0 A_{-x} \rangle$. We convert the integral into a contour integral using $\tilde{z}=i e^{ik}$ again and obtain
\begin{eqnarray}
\langle B_0 A_{\pm x} \rangle&=&\text{Re}\left\{\frac{ i^x \text{sgn}(J)}{\pi}\oint d\tilde z \tilde z^{|x|-1} \left[ \frac{4(J' \mp J)\pm \frac{8J/\gamma +\tilde z}{1+\tilde z^2}\gamma }{4J'}\right] \tilde F(\tilde z)^{1/2} \right\}, \label{eq_BA2_nonzeroJ}\\
\tilde F(\tilde z) &=& \frac{16(J'^2 - J^2)}{\gamma^2} \tilde z^2 - \frac{8J}{\gamma} \tilde z + \left[\frac{32(J'^2+J^2)}{\gamma^2}-1\right]  +\frac{8J}{\gamma} \tilde z^{-1} + \frac{16(J'^2-J^2)}{\gamma^2}\tilde z^{-2},
\label{eq_BA3_nonzeroJ}
\end{eqnarray}
where the contour is along the upper arc of the unit circle in the counter-clockwise sense [solid line in Fig.~\ref{fig_branchcut}(b)]. Similar to the $J=0$ case, we deform the contour along the branch cut defined by the roots of $\tilde F(\tilde z)=0$ and the origin. Note that only the segment of the line integral along the branch cut is real and contributes to Eq.~(\ref{eq_BA1_nonzeroJ}). There are 4 and 2 real roots when $|J'|\neq |J|$ and $|J'|=|J|$, respectively. The correlation length is determined by the largest root with $|\tilde z_L| < 1$ through $\xi = -1/(2 \ln  |\tilde z_L|)$. For the purpose of obtaining the critical exponent, we solve $\tilde F(\tilde z)=0$ near the critical point and get $|\tilde z_L| \approx 1- (1/8-|J'|/\gamma)\gamma/|J|$. Therefore, for $J \neq 0$, we have
\begin{eqnarray}
\xi = \frac{|J|/\gamma}{2 (\frac{1}{8}-\frac{|J'|}{\gamma}) },
\end{eqnarray}
so $\nu = 1$.

Following the same procedure, the asymptotic behavior of the $z$-component correlation can be evaluated in a lengthy but straightforward way by expanding Eq.~(\ref{eq_BA2_nonzeroJ}). We shall not repeat the derivations but summarize the result of large and even $x$ expansion here:
\begin{eqnarray}
C^{zz}(|x|) &=&\left\{
\begin{array}{ll}
 \frac{4(J^2-J'^2)}{\pi} |x|^{-3}\left\{\left[\sum_{i=1,2}\tilde z_i^{|x|}\left( g_{i,0}^+ +\frac{3 g_{i,1}^+}{2|x|}\right)\right]\left[\sum_{i=1,2}\tilde z_i^{|x|}\left( g_{i,0}^- +\frac{3 g_{i,1}^-}{2|x|}\right)\right] + \mathcal O \left( \frac{1}{x^2}\right)\right\},& |J'|>|J|\\ & \\
 \frac{4(J'^2-J^2)}{\pi} \tilde z_2^{2|x|} |x|^{-3} \left[\left( g_{2,0}^+ +\frac{3 g_{2,1}^+}{2|x|}\right)\left( g_{2,0}^- +\frac{3 g_{2,1}^-}{2|x|}\right) + \mathcal O \left( \frac{1}{x^2}\right)\right], & |J'|<|J|\quad\quad\quad\\ & \\
 \frac{2 J}{\pi} \tilde z_1^{2|x|} |x|^{-3} \left[\left( g_{1,0}^+ +\frac{3 g_{1,1}^+}{2|x|}\right)\left( g_{1,0}^- +\frac{3 g_{1,1}^-}{2|x|}\right) + \mathcal O \left( \frac{1}{x^2}\right)\right], & |J'| = |J|
\end{array}\right.
\end{eqnarray}
where the coefficients $g_{i,n}^\pm$ are obtained from the expansion:
\begin{eqnarray}
\label{eq_Gexpansion}
G^\pm(\tilde z_i y) &=& \sum_{n=0}g_{i,n}^\pm (1-y)^{\frac{2n+1}{2}},
\end{eqnarray}
where we have defined
\begin{eqnarray}
G^\pm(\tilde z ) &=& \left\{
\begin{array}{ll}
 \left[ \frac{4(J' \mp J)\pm \frac{8J/\gamma +\tilde z}{1+\tilde z^2}\gamma }{4J'}\right] \sqrt{\frac{(\tilde z-\tilde z_1)(\tilde z-\tilde z_2)(\tilde z-\tilde z_3)(\tilde z-\tilde z_4)}{\tilde z^2}}, & |J'|\neq|J|\\ & \\
 \left[ \frac{4(J' \mp J)\pm \frac{8J/\gamma +\tilde z}{1+\tilde z^2}\gamma }{4J'}\right] \sqrt{\frac{(\tilde z-\tilde z_1)(\tilde z-\tilde z_2)}{\tilde z}}. & |J'| = |J|
\end{array}\right. \nonumber
\end{eqnarray}
The roots of Eq.~(\ref{eq_BA3_nonzeroJ}) are labeled such that $|\tilde z_1| < |\tilde z_2| < 1 < |\tilde z_3| < |\tilde z_4| $ when $|J'| \neq |J|$, and $|\tilde z_1| < 1 < |\tilde z_2| $ when $|J'| = |J|$. Fig.~\ref{fig_branchcut}(b) provides a particular example of the branch structure. We note that the purpose of the expansion in Eq.~(\ref{eq_Gexpansion}), though not convergent around $y=1$, is to obtain the first few coefficients $g_{i,n}^\pm$ for the correlation function, according to Ref.~\cite{wu61}.


\twocolumngrid

\bibliography{herald}

\end{document}